\documentclass[aps,prl,twocolumn,superscriptaddress,showpacs,floatfix,reprint,longbibliography]{revtex4-1}

\usepackage{epsfig,amsmath,amssymb,latexsym}

\usepackage{graphicx}
\usepackage{subfigure}
\usepackage[normalem]{ulem}
\usepackage{soul}
\usepackage{color}
\usepackage[colorlinks,
            linkcolor=blue,
            anchorcolor=blue,
            citecolor=blue,
            urlcolor=blue
            ]{hyperref}
\definecolor{red}{rgb}{1,0,0}
\definecolor{blue}{rgb}{0,0,1}
\definecolor{black}{rgb}{0,0,0}

\begin{document}

\title{Ferromagnetism in a Semiconductor with Mobile Carriers via Low-Level Nonmagnetic Doping} \preprint{1}

\author{Bing-Hua Lei}
 \affiliation{Department of Physics and Astronomy, University of Missouri, Columbia, Missouri 65211-7010, USA}
\author{David J. Singh}
 \email{singhdj@missouri.edu}
 \affiliation{Department of Physics and Astronomy, University of Missouri, Columbia, Missouri 65211-7010, USA}
 \affiliation{Department of Chemistry, University of Missouri, Columbia, MO 65211, USA}

\date{\today}

\begin{abstract}
We show that doped cubic iron pyrite, which is a diamagnetic semiconductor,
becomes ferromagnetic when $p$-type doped.
We furthermore find that this material can exhibit high spin polarization both for tunneling and
transport devices.
These results are based on first principles electronic structure and transport calculations.
This illustrates the use of $p$-type doping without magnetic impurities as a strategy for obtaining ferromagnetic
semiconducting behavior, with implications for spintronic applications that require
both magnetic ordering and good mobility. This is a combination that has been difficult to achieve by
doping semiconductors with magnetic impurities.
We show that phosphorus and arsenic may be effective dopants for achieving this behavior.
\end{abstract}

\maketitle

\section{Introduction}

The search for ferromagnetic semiconductors was very active some twenty years ago
and continues to be of considerable interest.
\cite{Dietl_2002,dietl}
This is motivated in
part by potential uses in enabling spintronic technologies,
\cite{Wolf2001,Zutic2004}
and in part by excitement caused by
discoveries such as the ferromagnetic behavior of GaAs when heavily doped by Mn.
\cite{ohno}
This system has the particular advantages of being cubic and being based on a well-known
useful semiconductor material, particularly in the context of optoelectronics.
\cite{yoon}
In this system, replacement of trivalent Ga by Mn$^{2+}$ ions
introduces local moments on the Mn as well as $p$-type doping.
The induced carriers are spin-polarized
by interaction with the Mn moments, and mediate the coupling between them, yielding ferromagnetism.
However, the strong interaction of the carriers with disordered Mn$^{2+}$ on the lattice
invariably leads to strong scattering,
low mobility and loss of the good semiconducting properties of conventionally
doped non-magnetic GaAs.

Thus finding ways of achieving ferromagnetism in a doped semiconductor, while retaining
mobility, remains a challenge to be addressed.
In this regard, the use of itinerant magnetism rather than ordering of local moments offers a potential
path forward.
The itinerant Stoner model \cite{Stoner1939,Andersen_1977,Krasko_1987} predicts a magnetic instability
of the paramagnetic state
when the quantity $N(E_F)I\geq1$,
where $N(E_F)$ is the Fermi level, $E_F$, electronic density of states (DOS)
on a per atom basis for one spin and $I$ is an interaction parameter
that is material dependent, but typically $\sim$ 1 eV for 3$d$ transition metal elements.
\cite{Janak1977}
In this approach a Stoner instability caused by a high $N(E_F)$ leads to
magnetism.
However, high $N(E_F)$ is generally not compatible with semiconducting behavior, since it implies either
very high metallic carrier concentrations or very heavy effective mass.

We note that a similar problem occurs in thermoelectric materials where high
conductivity and high thermopower are both desired but high thermopower requires
high effective mass and high conductivity requires low effective mass.
\cite{He_2017,gaultois,yang}
This may be resolved by observing that the relevant effective masses for the conductivity and thermopower
can be different in materials that do not follow the single parabolic band model.
\cite{xing_2017,feng}
In the context of magnetism, this concept was used to propose that marcasite structure FeAs$_2$ may
be near a doping induced magnetic quantum critical point.  \cite{Lei_2020}
However, marcasite FeAs$_2$ has a non-cubic structure with a small band gap and while perhaps useful
as a thermoelectric, it is unsuitable for normal semiconducting applications.
\cite{Fan1972}

Importantly, both itinerant and strongly correlated electron models predict metallic ferromagnetism
near half-filling, at low carrier concentrations, and under other conditions, for example with flat bands.
\cite{linden,mielke}
Recently, this has been experimentally demonstrated in twisted bilayer graphene.
\cite{sharpe}
The two dimensional (2D) tJ model, which is a high $U$ limit of the Hubbard model and has
a high density of states at the band edge in 2D can be either ferromagnetic or ferrimagnetic
and this is also thought to be the case, although in a limited range of carrier concentration, for the Hubbard model.
Na$_x$CoO$_2$ and its hydrated counterpart may be examples of correlated materials near ferromagnetism
for this reason.
\cite{boothroyd,bayrakci,merino,cao}
It is also noteworthy that Na$_x$CoO$_2$ is also an excellent oxide thermoelectric.
\cite{terasaki}

Here we show that cubic pyrite $\beta$-FeS$_2$, which is a common mineral and a known semiconductor,
thought to be suitable for applications,
\cite{Ennaoui_1993,Cab_n_Acevedo_2014,Gong_2013,Wang_2014,Qiu_2013,Samad_2015},
particularly optoelectronic devices such as solar cells,
\cite{Cab_n_Acevedo_2012,Berry_2012,Shukla_2014,Rahman_2020,Wu_2016}
can be made magnetic by suitable $p$-type doping and that it has properties consistent with spintronic
applications.
Importantly, although FeS$_2$ contains a magnetic element, it is diamagnetic, meaning that
it does not contain Fe moments in its ordinary state. The magnetism that we find is itinerant
in nature. It is not the consequence of introduction of local moments into the $p$-type doped material.
We note that magnetism also was recently shown to occur in electrostatically
gated $n$-type FeS$_2$ analogous to
the known ferromagnetic alloy (Fe,Co)S$_2$ and CoS$_2$,
\cite{jarrett,Miyahara_1968,ohsawa,mazin-CoS2}
although at very high metallic carrier concentrations that may not be suitable
for semiconducting applications.
\cite{walter,day-roberts}
Importantly,
FeS$_2$, although mainly studied with $n$-type doping, is synthesizable as $p$-type and does show
semiconducting behavior in this form.
\cite{lichtenberger,lehner,willeke,blenk}

The results here are obtained with standard density functional theory (DFT),
specifically the generalized gradient approximation (GGA) of Perdew, Burke and Ernzerhof.
\cite{Perdew1996}
It is generally possible to obtain magnetism in Fe compounds by the use of additional
terms in the Hamiltonian, such as in the LDA+$U$ method,
\cite{dudarev}
but we use the conservative PBE functional only.
In this regard, it should also be noted that the magnetic behavior of CoS$_2$ and the
alloy (Fe,Co)S$_2$ are well described by standard density functional calculations.
\cite{mazin-CoS2}

\section{First Principles Calculations}

The main calculations were done with the general potential linearized augmented plane-wave (LAPW) method
\cite{LAPW}
as implemented in the \textsc{WIEN2k} code.
\cite{WIEN2k}
We used well converged basis sets with the
convergence criterion $R_{min}k_{max}$=9, where $R_{min}$ is the smallest LAPW sphere radius
and $k_{max}$ is the planewave sector cutoff. We included local orbitals for the semicore states.
We use the cubic structure of FeS$_2$, space group $Pa\overline{3}$ as shown in Fig. \ref{fig:fig1},
lattice parameter $a$=5.416 \AA, from experiment
\cite{Rieder_2007}
with internal atomic coordinates relaxed.
Calculations were done both in a scalar relativistic approximation and with the inclusion of
spin orbit. In the present case spin orbit effects are small, as seen in the comparison of
the density of states given in Fig. \ref{fig:fig1}.
The scalar relativistic band gap of 0.390 eV is reduced only slightly to 0.387 eV when spin orbit
is included. Note, however, that band gaps are underestimated in standard GGA calculations.
LAPW sphere radii of 2.0 bohr were used for both Fe and S.
Most calculations were done with Brillouin zone samplings consisting of 24x24x24 {\bf k}-point grids.
Tests were done with denser meshes, which were also used for the transport integrals.
These were done using the BoltzTraP code.
\cite{Madsen2006a}
Structure relaxations for the dopant containing supercells were done using the VASP code,
\cite{PAW,VASP}
with standard
settings and checked by verifying the the forces on atoms were small in LAPW calculations.

\section{Results and Discussion}

\begin{figure}[htbp]
 \centering
 \includegraphics[width=\columnwidth]{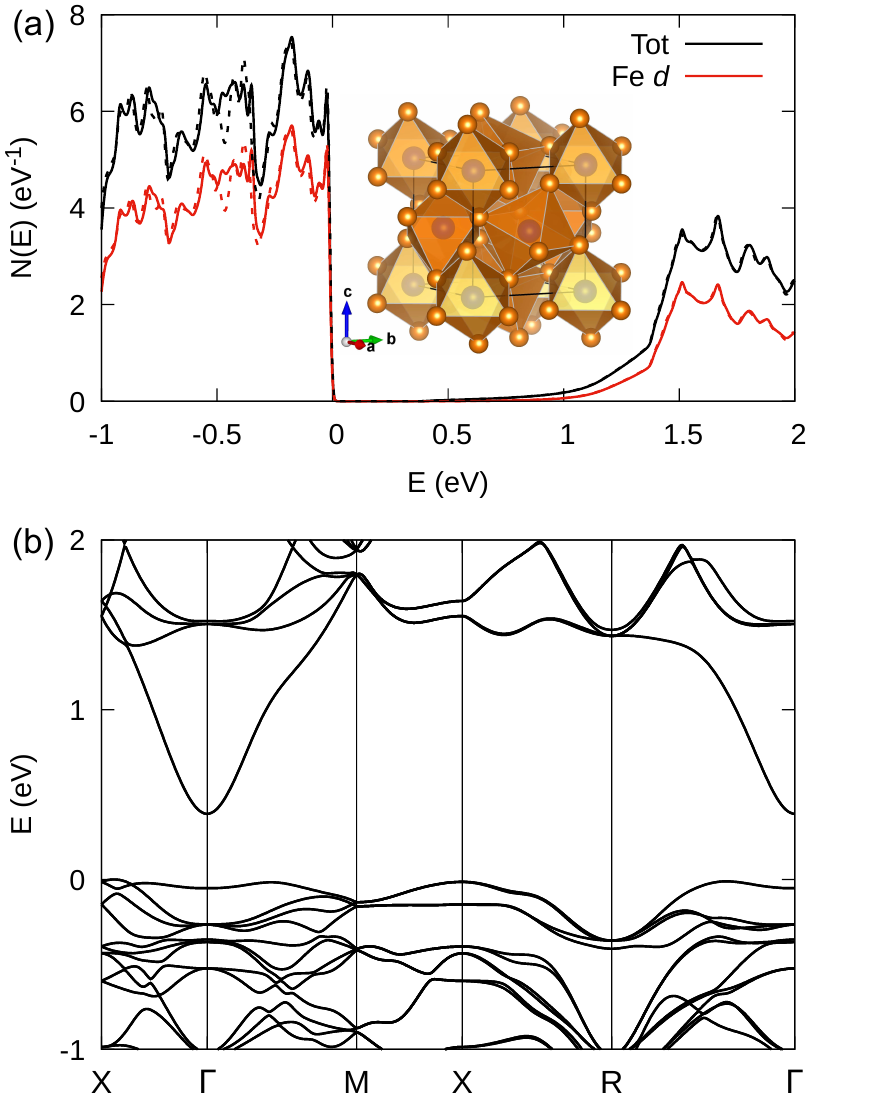}
 \caption{\label{fig:fig1}Electronic and crystal structure of FeS$_2$.
(a) Electronic density of states of FeS$_2$ per formula unit both spins and Fe $d$
projection, with a depiction of the crystal structure inset, showing
S-S bonds. The solid line is scalar relativistic and the dashed lines are
including spin orbit; (b) band structure with spin-orbit. The energy zero is the valence band maximum
and the projection is of d character within the Fe LAPW sphere of radius 2.0 bohr.}
\end{figure}

FeS$_2$ has a centrosymmetric cubic crystal structure, space group $Pa\overline{3}$, with four formula
units per primitive cell (inset of Fig. \ref{fig:fig1}a).
Important features of the crystal structure are the octahedral coordination of Fe and the S-S bonding.
This S-S bonding
leads to a splitting of the S bands into bonding, non-bonding and antibonding components.
The lower part
of the conduction bands comes from antibonding S states that overlap the upper Fe $e_g$
crystal field level.
The valence bands come from Fe $t_{2g}$ orbitals, S derived bands below these.
As discussed previously,
\cite{mazin-CoS2}
this leads to an Fe$^{2+}$ valence state with six $d$-electrons, that occupy $t_{2g}$ crystal field level.
This is seen in the band structure (Fig. \ref{fig:fig1}b), where the valence bands are very narrow, with a calculated $t_{2g}$ width of
1.27 eV. This narrow width is consistent with photoemission data.
\cite{lichtenberger}

The calculated band structure in Fig. \ref{fig:fig1} shows
an indirect band gap of 0.39 eV.
Previous calculations show a wide range of gaps depending on the method used.
\cite{Banjara_2018,Hu_2012,Opahle_1999,Muscat_2002}
It is to be noted that standard PBE calculations such as those reported here typically underestimate band gaps.
Experimental values for FeS$_2$ range from 0.73 to 1.2 eV 
\cite{Banjara_2018,Hu_2012,Abass_1987,Karguppikar_1988,Tsay_1993,Kou_1978,Yang_1995,Seehra_1979}
and an indirect band gap of 0.95 eV is frequently quoted.
\cite{Ennaoui_1984,Schlegel_1976}

The result is that the DOS onset is much sharper for $p$-type than for $n$-type.
This is also evident in the band structure. The conduction band edge is from a single dispersive band
at the $\Gamma$ point associated with antibonding combinations of S $p$ orbitals.
The valence band edge is from several flatter bands.

Furthermore, there is an extremely sharp onset of the valence band DOS at the band edge, 
unlike the usual isotropic parabolic band square root behavior common in cubic semiconductors.
The reason is the very flat band behavior along some directions, e.g. $\Gamma$-$X$, with more dispersive
behavior in the top valence band in other directions.
In some directions the bands have significant dispersion as seen near $R$,
and in other directions there is a mixture of heavy and light bands as seen along $\Gamma$-$X$.

This highly anisotropic behavior in a cubic material is reminiscent of the behavior
seen in several good thermoelectrics including $p$-type PbTe,
\cite{parker} and $n$-type SrTiO$_3$, \cite{sun}
where it leads to an effective reduction in the dimensionality and sharper onset of the DOS
similar to what is found here.
The resulting decoupling of DOS and transport effective mass is important
for the thermoelectric performance.
Here the consequence is that
the Stoner criterion
is exceeded for $E_F$ very close to the band edge, and therefore is expected to be reachable by $p$-type doping,
at very much lower carrier concentrations than for $n$-type.
This is the case even though $p$-type FeS$_2$ is a semiconductor.

We did virtual crystal calculations with constrained density functional theory
employing the fixed spin moment procedure to investigate magnetism in $p$-type FeS$_2$,
as well as self consistent virtual crystal calculations.
The virtual crystal approximation is a self-consistent average potential
method that does not rely on rigid bands and is appropriate for doping provided that the potential
due to the dopant atoms is not too strong.
In the present calculations virtual crystal doping was done by changing the nuclear charge on the S site
and adjusting the number of electrons accordingly.

\begin{figure}[htbp]
 \centering
 \includegraphics[width=0.94\columnwidth]{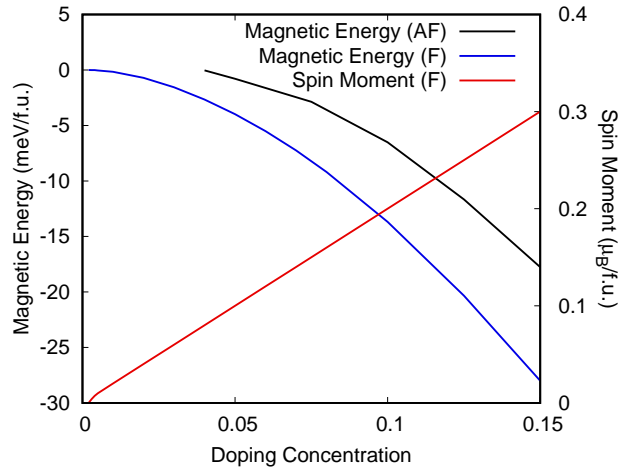}
 \caption{\label{fig:fig2}Magnetism in p-type FeS$_2$.
Scalar relativistic ferromagnetic spin moment and magnetic energy per
formula unit of $p$-type virtual crystal
doped FeS$_2$. The concentration is the change in charge on the S site. Since there are two
S per formula unit, the hole concentration is twice the doping concentration on a per formula
unit basis. The magnetic energy is calculated for ferromagnetic (F) and antiferromagnetic
ordering (AF, see text).}
\end{figure}

The results are shown in Fig. \ref{fig:fig2}.
As seen, itinerant ferromagnetism starts at very low doing.
We find that the onset of ferromagnetism is at a doping concentration of 0.003 on the S site
(0.006 holes per formula unit).
Furthermore, the holes become fully spin polarized to yield half metallic behavior at a doping
concentration of 0.005.
Above this concentration the spin moment is equal to the hole concentration.
As seen, the magnetic energy is approximately quadratic in the moment in this regime.
We also did self-consistent calculations including spin-orbit.
These yield similar results. Specifically, the magnetic energies are little changed,
spin moments are reduced by $\sim$2\% and an orbital moment on Fe of $\sim$10\% of the spin moment
is induced. This orbital moment is parallel to the spin moment, consistent with Hund's third rule for
a $d^6$ system.

We also investigated the possibility of antiferromagnetism. This was done for antiferromagnetic orderings
of the four Fe atoms within a unit cell, with alternation of spin directions along a Cartesian direction.
With this arrangement each Fe has eight antiferromagnetic bonds and four ferromagnetic bonds, typical of
collinear antiferromagnets on an fcc lattice, such as the Fe sublattice of FeS$_2$.
The magnetic energies of these antiferromagnetic arrangements are shown in Fig. \ref{fig:fig2}.
The AF solution is always higher in energy than the ferromagnetic, and therefore disfavored, over the
entire doping range considered.

\begin{figure}[htbp]
\centering
 \includegraphics[width=\columnwidth]{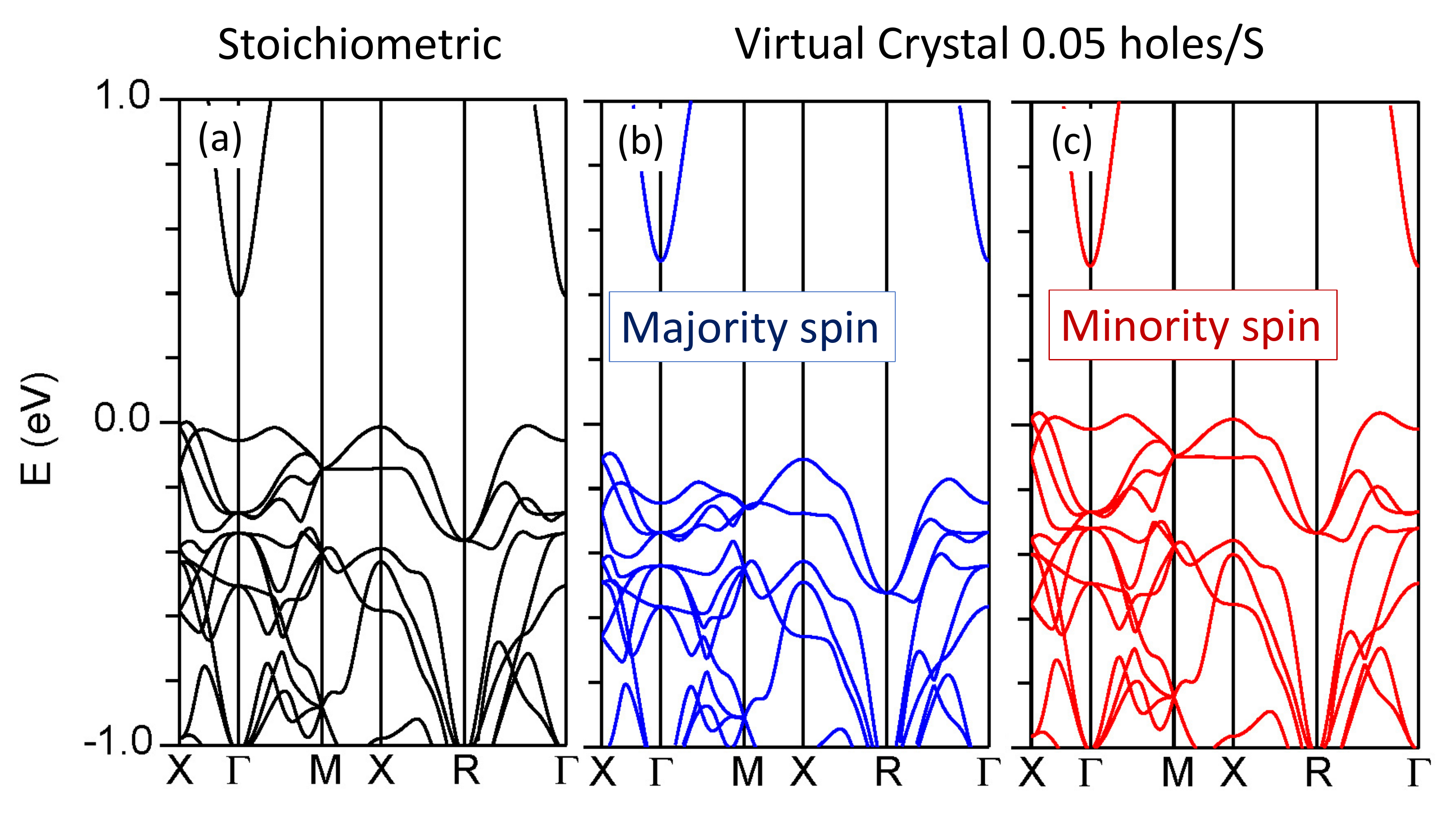}
\caption{\label{fig:bands-compare}Comparison of scalar relativistic band structures for pristine FeS$_2$
(a)
and ferromagnetic FeS$_2$ with a virtual crystal doping concentration of 0.05 on the S site (0.1 holes
per Fe), showing majority and minority spin bands in (b) and (c), respectively.
Note the very similar band structures.}
\end{figure}

The band structure of pristine FeS$_2$ is compared with the ferromagnetic
virtual crystal band structure for an electron deficiency of 0.05 per S in Fig. \ref{fig:bands-compare}.
As seen, the position of the valence band maximum and band shapes near the valence band maximum are
practically the same for the three cases, aside from the exchange splitting.
This behavior supports the use of the virtual crystal approximation.
The calculated exchange splitting at the valence band maximum for this doping level is 0.128 eV.
In contrast, the exchange splitting at the conduction band minimum is very small reflecting the
S character of the states making up this band.
In any case, with this exchange splitting of the valence band
and the fully half-metallic predicted state, the transport is expected
to be fully spin polarized. This is favorable for mobility as spin-flip scattering is suppressed in a half-metal.
Specifically, in the fully polarized case, only minority spin carriers can contribute to conduction,
and scattering into majority spin states is suppressed by a lack of available states within
$k_BT$ of the Fermi level.
This half-metallic behavior is also particularly interesting from the point of view of devices,
since it implies fully polarized transport.
This can lead to strong negative magnetoresistance, and applications based on injection of spin-polarized
carriers and spin-valves.
\cite{ohno2,katsnelson,pickett}

We now turn to the question of chemical doping.
In general doping at sufficient levels and with sufficient uniformity to achieve ferromagnetism
in semiconductors is a considerable challenge and problems such as aggregation of impurities
or surface segregation have led to controversial results.
\cite{dietl2,coey}
Naturally occurring FeS$_2$ (the mineral pyrite)
typically contains various impurities as well as S vacancies, although the S/Fe ratio is generally
very close to 2.
\cite{Blanchard_2007,Abraitis_2004} 
Prior DFT  work shows indicates that As can enter pyrite 
via a substitution mechanism where arsenic substitutes for sulfur in the disulphide group,
resulting in (AsS)$^{3-}$ dianions replacing S$_2^{2-}$ in the lattice,
\cite{Blanchard_2007} and that furthermore sizable mole fractions may be achievable.
\cite{Reich_2006}
However, it should be noted that there are competing phases such as FeAsS.

Experimentally, $p$-type doing via As has been demonstrated with carrier concentrations up to the
10$^{18}$ cm$^{-3}$ range although with a low mobility of $\sim$ 8 cm$^2$V$^{-1}$s$^{-1}$ at room
temperature and evidence for multiple As sites in the material as grown.
\cite{lehner}
In any case, As may be a useful $p$-type dopant if samples can be improved.
This is supported by theoretical calculations, \cite{lehner}
which also suggest P as a dopant. \cite{Hu_2012}
This is consistent with experimental reports of $p$-type transport in P doped FeS$_2$ crystals.
\cite{willeke,blenk}
Hu and co-workers note that the group V impurities on the S are spin polarized.
\cite{Hu_2012}
This is a consequence of the odd number of electrons. However, as we find ferromagnetism can occur
at low levels of doping, which is not a consequence of the electron count per impurity.

An important issue in developing ferromagnetic semiconducting $p$-type FeS$_2$ will be achieving
suitable doping levels, while maintaining sample quality so that carriers remain mobile.
One favorable aspect in this regard is the reasonably high dielectric constant, $\epsilon_0$=10.9
\cite{husk}
similar to semiconductors such as GaAs ($\epsilon_0$=12.9).
High dielectric constant in general
disfavors Anderson localization, which may be an important issue here due to high band mass,
and is also beneficial for reducing ionized impurity scattering.
\cite{mao,he}

\begin{figure}[htbp]
 \centering
 \includegraphics[width=0.95\columnwidth]{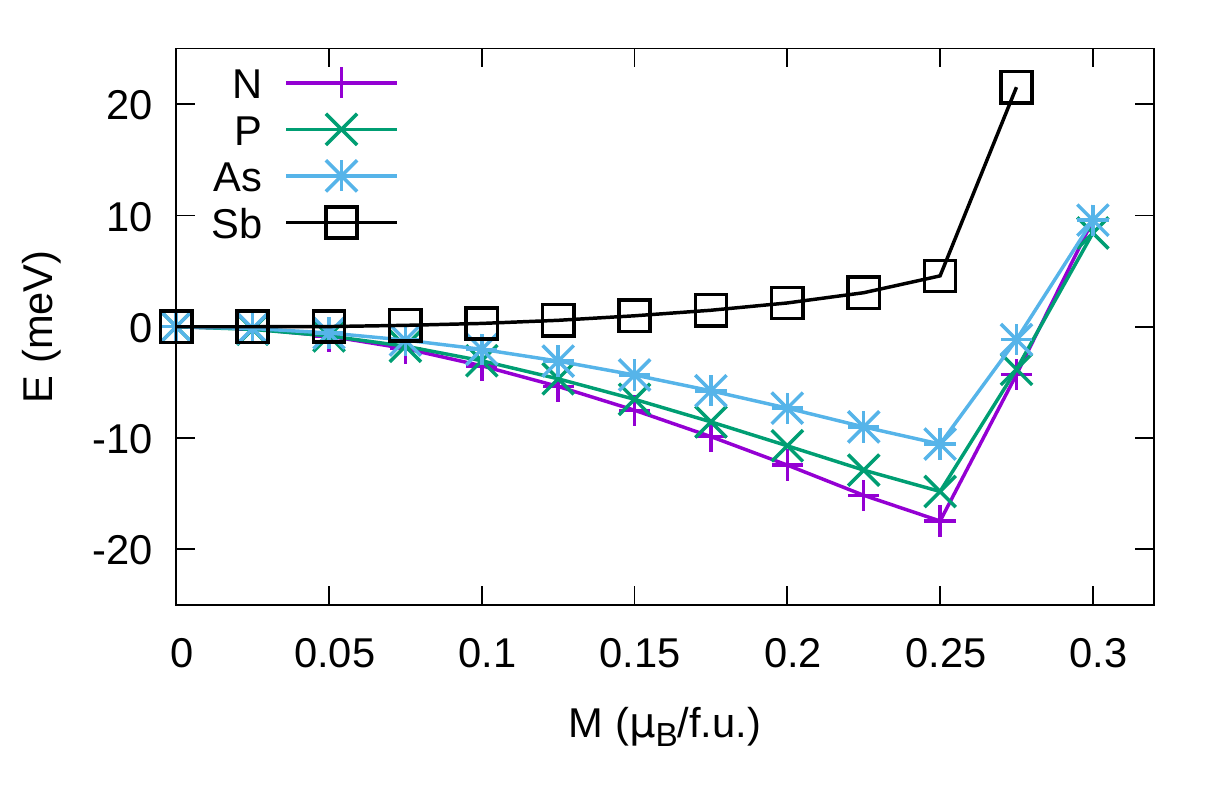}
 \caption{\label{fig:fig3}Magnetism in doped supercells of FeS$_2$.
Fixed spin moment energy of FeS$_2$ doped by
substitution of group five group for S at a level of 12.5\% based on supercell calculations.}
\end{figure}

We constructed supercells for a doping level of 12.5\% for various group five elements including As,
and searched for magnetism using fixed spin moment calculations.
These cells have one S atom substituted and were fully relaxed, including
lattice parameters and internal coordinates, using the VASP code \cite{PAW,VASP}
prior to the calculation of the magnetic properties.
Results are shown in Fig. \ref{fig:fig3}.
Magnetism is clearly found for N, P and As.

We note that N forms very stable binary phases with transition metals, and may not be a practical dopant.
However, the results for P and As suggest these elements as dopants.
The induced holes are fully spin polarized in the DOS.
This is consistent with the fixed spin moment calculations, which show the minimum energy at the
maximum moment of 0.25 $\mu_B$ that can be obtained based on the hole count of 0.25 holes per formula unit.

The above results show that it is possible to obtain magnetism in $p$-type FeS$_2$.
However, this is not sufficient for a magnetic semiconductor.
Transport properties are also important.
In a 3D parabolic band semiconductor,
the density of states is $N(E_F$) $\sim$ $m^*$ E$_F^{1/2}$, with E$_F$ relative to the band edge,
while the mobility generally varies as $\mu \sim 1/m^*$.
Additionally, the Curie temperature in itinerant models generally decreases with the decreases in the energy scale,
which is governed by 1/$m^*$.
Thus high $m^*$ is not desirable in a magnetic semiconductor, but is essential for the Stoner
mechanism discussed above.

This can be resolved by band structures with multiple bands at the band edge and non-parabolic band
dispersions, for example materials where there are both dispersive and flat bands near the band edge.
\cite{Lei_2020}
Importantly, the band structure of FeS$_2$ shows
that besides flat bands near the valence band edge that lead to the high DOS onset,
there are also more dispersive bands, in particular near the $R$ point.
In addition decoupling of the DOS and transport mass is possible due to carrier
pocket anisotropy when the band extremum is away from the highest symmetry points ($\Gamma$ and $R$) as
is the case here.
This has been discussed previously in the context of thermoelectrics.
\cite{Parker_2015,witkoske}

We used the BoltzTraP code \cite{Madsen2006a} to evaluate the transport effective mass.
This is the effective mass that in a parabolic band model would give the same conductivity as the actual
band structure for a given scattering time in Boltzmann transport theory.
This in general depends on both temperature and carrier concentration.
Including spin-orbit, the calculated transport effective mass is 8.0 $m_e$ for $p$-type and
0.41 $m_e$ for $n$-type.
It is remarkable that even though there is a sharply increasing DOS near the valence band edge,
normally not consistent with semiconducting behavior in a cubic material,
the transport effective mass is not so high as to preclude $p$-type semiconducting behavior.
This is also consistent with experimental data.
For example, $p$-type films grown by reactive magnetron sputtering,
showed room temperature Hall mobility of 25 cm$^2$V$^{-1}$s$^{-1}$
even with a high carrier concentration of $\sim$10$^{20}$ cm$^{-3}$.
\cite{lichtenberger}
This carrier concentration (10$^{20}$ cm$^{-3}$) corresponds to a S doping fraction of 0.002,
which is not so far from the onset of ferromagnetism at a doping fraction of $\sim$0.003 according to
the present calculations.

\begin{figure}[htbp]
 \centering
 \includegraphics[width=0.9\columnwidth]{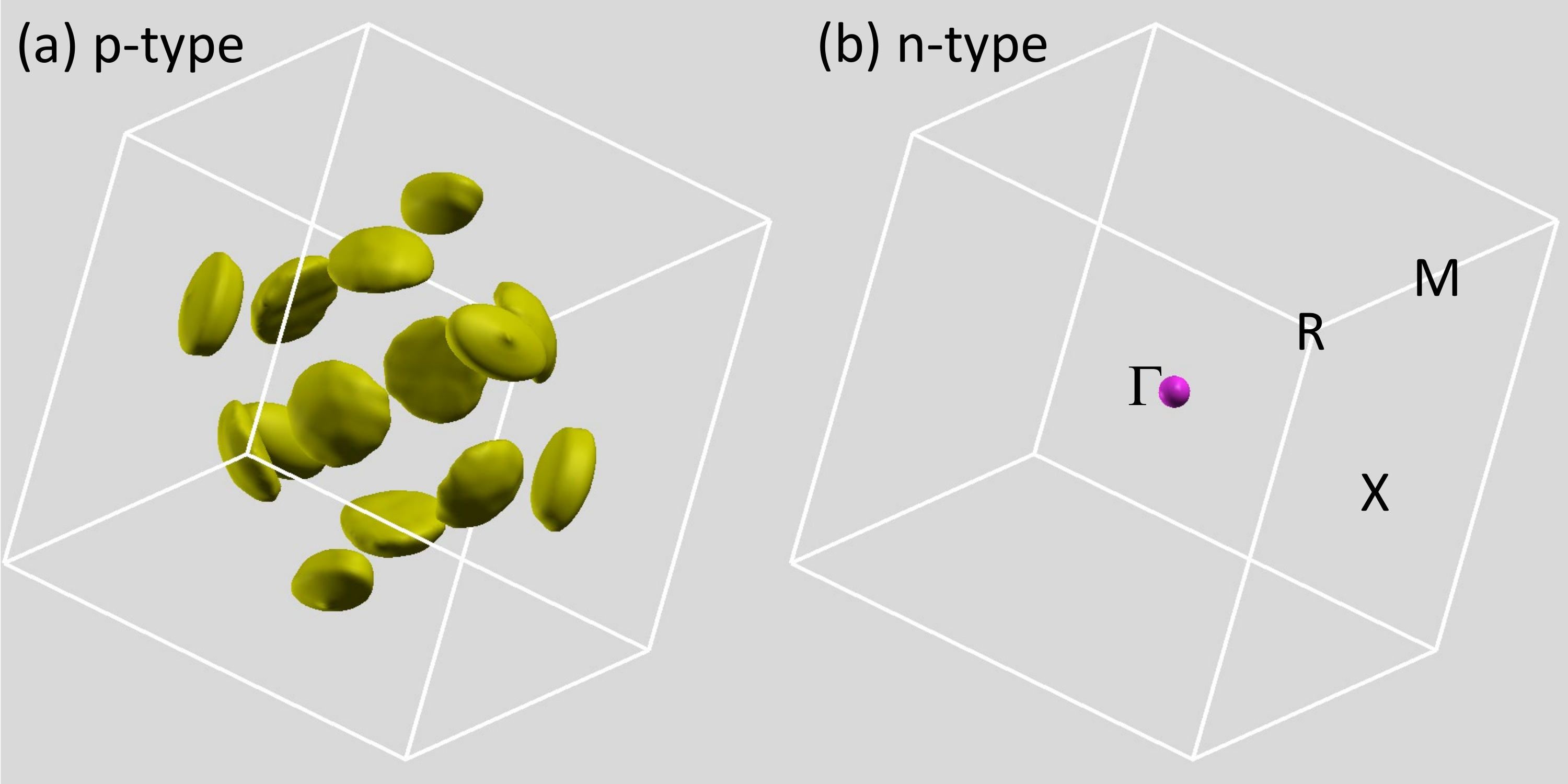}
\caption{\label{fig:fig6}Carrier pocket visualization for FeS$_2$, including spin-orbit. (a) $p$-type,
0.015 eV below the valence band maximum and (b) $n$-type 0.015 eV above the conduction band minimum.
Note the single nearly spherical conduction band pocket and the multiple highly anisotropic valence
band pockets.}
\end{figure}

Fig. \ref{fig:fig6} visualizes the valence and conduction band carrier pockets for FeS$_2$.
The conduction band edge has a single carrier pocket at $\Gamma$ with a spherical shape
as expected for a single parabolic band in a cubic material.
In contrast, the valence band shows a high carrier pocket degeneracy from band maxima along
$\Gamma$-$X$ and $\Gamma$-$R$. These carrier pockets are very anisotropic taking pancake-like shapes.
This is a characteristic shape that leads to a lighter transport effective mass than would
be inferred from the DOS, as was discussed in the context of thermoelectrics.
\cite{Parker_2015}

\section{Summary and Conclusions}

In summary, we illustrate the feasibility of an itinerant method for obtaining ferromagnetic semiconductors.
As mentioned, we obtain full spin polarization with As and P doping.
This is
with moderate carrier concentrations and high spin polarization by the example of $p$-type FeS$_2$.
The key is to break the connection between transport and density of states effective mass using 
complex band structures in analogy with thermoelectric materials.
We find that $p$-type FeS$_2$ may be such a material and that doping may be achievable using As and/or P
substitution on the S sites.
This shows a surprising connection between thermoelectrics and magnetic semiconductors,
which may lead to further discoveries of novel magnetic semiconducting materials.

\section{Acknowledgments}

This work was supported by the U.S. Department of Energy, Basic Energy Sciences, Award Number DE-SC0019114.

\bibliography{References}

\end{document}